 [2005/06/22 5.2/AAS markup document class]%
\documentclass[iop]{emulateapj}

\makeatletter % This code stolen from ltxguide.cls:

\let\o@verbatim\verbatim
\def\verbatim{%
  \ifhmode\unskip\par\fi
  \ifx\@currsize\normalsize
     \small
  \fi
  \o@verbatim
}

\renewcommand \verbatim@font {%
  \normalfont \ttfamily
  \catcode`\<=\active
  \catcode`\>=\active
}

\RequirePackage{shortvrb}
\MakeShortVerb{\|}

\begingroup
  \catcode`\<=\active
  \catcode`\>=\active
  \gdef<{\@ifnextchar<\@lt\@meta}
  \gdef>{\@ifnextchar>\@gt\@gtr@err}
  \gdef\@meta#1>{\m{#1}}
  \gdef\@lt<{\char`\<}
  \gdef\@gt>{\char`\>}
\endgroup
\def\@gtr@err{%
   \ClassError{ltxguide}{%
      Isolated \protect>%
   }{%
      In this document class, \protect<...\protect>
      is used to indicate a parameter.\MessageBreak
      I've just found a \protect> on its own.
      Perhaps you meant to type \protect>\protect>?
   }%
}
\def\verbatim@nolig@list{\do\`\do\,\do\'\do\-}

\newcommand{\m}[1]{\mbox{$\langle$\it #1\/$\rangle$}}

\def\cmd#1{\cs{\expandafter\cmd@to@cs\string#1}}
\def\cmd@to@cs#1#2{\char\number`#2\relax}
\DeclareRobustCommand\cs[1]{\texttt{\char`\\#1}}
\def\GetFileInfo#1{%
  \def\filename{#1}%
  \def\@tempb##1 ##2 ##3\relax##4\relax{%
    \def\filedate{##1}%
    \def\fileversion{##2}%
    \def\fileinfo{##3}}%
  \edef\@tempa{\csname ver@#1\endcsname}%
  \expandafter\@tempb\@tempa\relax? ? \relax\relax}

\expandafter\GetFileInfo\expandafter{\jobname.tex}%

\begin{document}

\makeatother

\title{Scientific verification of Faraday Rotation Modulators:\\ Detection of diffuse polarized Galactic emission}

\author{S. Moyerman\altaffilmark{1}, E. Bierman\altaffilmark{1}, P. A. R. Ade\altaffilmark{2}, R. Aiken\altaffilmark{3}, D. Barkats\altaffilmark{4}, C.Bischoff\altaffilmark{10}, J. J. Bock\altaffilmark{5}, H. C. Chiang\altaffilmark{6}, C. D. Dowell\altaffilmark{5}, L. Duband\altaffilmark{7}, E. F. Hivon\altaffilmark{8}, W. L. Holzapfel\altaffilmark{9}, V. V. Hristov\altaffilmark{3}, W. C. Jones\altaffilmark{3,6}, J. Kaufman\altaffilmark{1}, B. G. Keating\altaffilmark{1}, J. M. Kovac\altaffilmark{10}, C. L. Kuo\altaffilmark{11,12}, E. M. Leitch\altaffilmark{13}, P. V. Mason\altaffilmark{3}, T. Matsumura\altaffilmark{14}, H. T. Nguyen\altaffilmark{5}, N. Ponthieu\altaffilmark{15}, C. Pryke\altaffilmark{16}, S. Richter\altaffilmark{3}, G. Rocha\altaffilmark{5}, C. Sheehy\altaffilmark{16}, Y. D. Takahashi\altaffilmark{9}, J. E. Tolan\altaffilmark{11,12}, E. Wollack\altaffilmark{17}, K. W. Yoon\altaffilmark{11}}
\email{smoyerma@ucsd.edu}

\altaffiltext{1}{Center for Astrophysics and Space Sciences, University of California,
    San Diego, CA 92037}
\altaffiltext{2}{Department of Physics and Astronomy, University of Wales, Cardiff, CF24 3YB, Wales, UK}
\altaffiltext{3}{Department of Physics, California Institute of Technology, Pasadena, CA 91125, USA}
\altaffiltext{4}{Joint ALMA Observatory, ESO, Santiago, Chile}
\altaffiltext{5}{Jet Propulsion Laboratory, Pasadena, CA 91109, USA}
\altaffiltext{6}{Department of Physics, Princeton University, Princeton, NJ 08544, USA}
\altaffiltext{7}{SBT, Commissariat a l'Energie Atomique (apostrophe), Grenoble, France}
\altaffiltext{8}{Insititut d'Astrophysique de Paris, Paris, France}
\altaffiltext{9}{Department of Physics, University of California at Berkeley, Berkeley,
CA 94720, USA}
\altaffiltext{10}{Harvard-Smithsonian Center for Astrophysics, 60 Garden Street, Cambridge, MA 02138, USA}
\altaffiltext{11}{Stanford University, Palo Alto, CA 94305, USA}
\altaffiltext{12}{Kavli Institute for Particle Astrophysics and Cosmology (KIPAC), Sand
Hill Road 2575, Menlo Park, CA 94025, USA}
\altaffiltext{13}{University of Chicago, Chicago, IL 60637, USA}
\altaffiltext{14}{High Energy Accelerator Research Organization (KEK), Tsukuba, Ibaraki, Japan}
\altaffiltext{15}{Insititut d'Astrophysique de Spaciale, Université Paris-Sud, Orsay, France}
\altaffiltext{16}{Department of Physics, University of Minnesota, 116 Church
Street S.E. Minneapolis, MN, USA 55455}
\altaffiltext{17}{Code 665, NASA/Goddard Space Flight Center, Greenbelt, MD 20771}

\begin{abstract}
The design and performance of a wide bandwidth linear polarization modulator based on the Faraday effect is described. Faraday Rotation Modulators (FRMs) are solid-state polarization switches that are capable of modulation up to ~10 kHz. Six FRMs were utilized during the 2006 observing season in the Background Imaging of Cosmic Extragalactic Polarization (BICEP) experiment; three FRMs were used at each of BICEP's 100 and 150 GHz frequency bands. The technology was verified through high signal-to-noise detection of Galactic polarization using two of the six FRMs during four observing runs in 2006. The features exhibit strong agreement with BICEP's measurements of the Galaxy using non-FRM pixels and with the Galactic polarization models. This marks the first detection of high signal-to-noise mm-wave celestial polarization using fast, active optical modulation. The performance of the FRMs during periods when they were not modulated was also analyzed and compared to results from BICEP's 43 pixels without FRMs. 

\end{abstract}

\section{Introduction}

Numerous experiments in the last two decades have shown that the cosmic microwave background (CMB) is a powerful cosmological probe. The temperature anisotropy of the CMB has now been mapped to exquisite precision by many experiments (e.g. - \cite{Jones06, Reichardt09, Friedman09, Sievers09, Jarosik10, Hlozek11, Reichardt11}), yielding tight  constraints on the standard $\Lambda$CDM cosmological model \citep{Brown09, Larson10, Dunkley10, Keisler11, Story12}. 

Many current and upcoming experiments focus on measuring CMB polarization, which potentially encodes information from long before the epoch of matter-radiation decoupling. CMB polarization can be decomposed into two orthogonal bases: the ``$E$-mode", a curl-free mode, and the ``$B$-mode", which possesses a curl or handedness. The shape of the $E$-mode power spectrum has been measured and is consistent with the $\Lambda$CDM cosmological model predicted by the temperature power spectrum. Multiple experiments have now measured the $E$-mode and temperature cross $E$-mode power spectrum \citep{Leitch05, Montroy06, Page07, Sievers07, Wu07, Bischoff08, Nolta08, Pryke09, Chiang2010, QUIET12}. 

Primordial $B$-mode polarization, however, is not a facet of the standard $\Lambda$CDM model of the universe. Instead, the inflationary paradigm predicts the existence of a stochastic gravitational wave background (GWB) that would imprint a unique pattern on the surface of last scattering \citep{Polnarev85}. While this faint signal does affect the CMB $E$-mode and temperature power spectra, the GWB's effect is degenerate with other parameters. Only the $B$-mode polarization provides a unique and direct probe of the GWB. This signal is expected to peak at degree angular scales with an amplitude determined by the energy scale of inflation \citep{Kamionkowski97}. A detection of the $B$-mode signal would provide strong evidence for an inflationary phase in the early Universe.

Detection of the sought-after $B$-mode signal presents many difficulties. First, the inflationary $B$-mode amplitude is extremely small compared to sky temperature, the CMB's temperature, and even its temperature and $E$-mode anisotropy. Currently, the most restrictive limits, derived from temperature anisotropy, put an upper limit on the expected $B$-mode signal of less than 200 nK \citep{Keisler11, Story12}. Second, polarized emission from galactic and extra-galactic objects acts as a foreground contaminant for CMB polarization measurements. Even in the case of optimistic predictions for the signal amplitude, models suggest that foreground contamination could be an order of magnitude larger than the $B$-mode signal at the largest angular scales \citep{Amblard07}. Finally, exquisite control of systematic and instrumental effects, down to the tens of nK level, will be required before a detection of $B$-modes can be claimed with confidence. 

One approach to mitigating systematic errors is to modulate incoming polarization, thereby shifting the signal to higher frequencies and away from lower frequency systematic contaminants. Rapid modulation eliminates concerns about time varying thermal, optical, or electrical fluctuations that often change on much longer time scales. In essence, modulation speed can be used to tune the signal band of the instrument, otherwise set by a combination of scan speed and beam size, placing the signal away from microphonic lines and low frequency $1/f$ noise. The lack of limitation on scan speed expands the parameter space for observation strategies. Modulation also mitigates polarized systematic effects that are introduced by optical elements between the modulator and the detector.

There are additional benefits to polarization modulation. The reconstruction of the Stokes Q (or U) parameter requires two independent measurements at different detector angles, typically done by two detectors. Mismatched detector properties, such as those caused by caused by differential spectral response, differential pointing, and mismatched transfer functions, can result in spurious polarization. Modulation allows a single linearly polarized detector to act as an independent polarimeter, removing systematics resulting from combining mismatched detectors.

Many CMB polarimeters already employ mechanisms to modulate the incident radiation field about the optical axis of the instrument. The modulator effectively exchanges electric fields between two detectors (or a single detector in two orientations) and decomposes the radiation into two orthogonal bases. Traditional methods for modulation in millimeter wave polarimetry include physical rotation of the entire instrument about the optical axis \citep{Halverson98, Keating01, Padin01, Yoon06, OSullivan08, Hinderks08, Ogburn10}, rotation or translation of a wiregrid polarizer \citep{Caderni78, Chuss12} or birefringent half-waveplate \citep{Murray92, Hildebrand00, Oxley05, Johnson07, EssingerHileman10, Filippini11, Keating11, Staniszewski12}, or parallactic angle rotation (``sky rotation") with respect to the fixed instrument coordinate system. 

These classical modes of polarization modulation often employ bulky and complex mechanisms, where any mechanical failure would result in the complete loss of polarization modulation. Faraday Rotation Modulators (FRMs) are coupled to individual detectors, eliminating the single point failure mode and relieving the need for producing large grids and waveplates that are often limited to tens of centimeters in diameter. FRMs can be tuned individually for specific bandwidths, whereas birefringent crystal waveplates require specialized anti-reflection coatings that are difficult to optimize for multiple frequencies. 

In addition, polarization modulation via mechanical motion is limited to a low frequency range due to  tolerances on vibrational and microphonic noise. Specifically, half-waveplate rotation is limited to modulation frequencies of tens of Hertz (Hz) at most and boresight rotation is limited to much less still. Parallactic angle rotation is also a slow signal modulation and is also location dependent; at South Pole observatories, such as BICEP, parallactic angle modulation caused by sky rotation is simply not present. 

In this paper we describe the Faraday Rotation Modulator, a solid-state polarization switch that employs the Faraday effect to rapidly modulate polarized light at cryogenic temperatures. The FRM is compact, works over a large frequency range, and is capable of modulation rates up to 10 kHz. Specifically, we describe the laboratory testing and Galactic observations of the FRMs in the BICEP experiment during the 2006 observing season. 

\section{FRM Design for Use in BICEP}

\begin{figure}[h]
  \centering
  \includegraphics[scale=0.3]{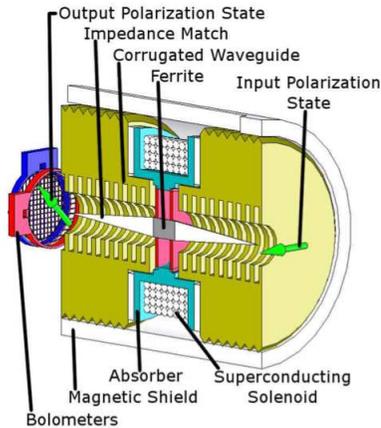}
  \caption[Cross Section of FRM]{Cross-Sectional schematic of a FRM. Polarized radiation traverses the FRM from right to left. The alumina cones serve as an impedance match and are attached to the ferrite in the center. The ferrite is supported by dielectric washers (shown in pink). A superconducting solenoid surrounds the ferrite and a corrugated waveguide surrounds the cones.}
\label{FRMschematic}
\end{figure}

Figure~\ref{FRMschematic} shows a cross-sectional schematic of a FRM. Along the path of incident radiation are two alumina cones attached to either side of a ferrite cylinder, providing an impedance match to minimize reflection loss. The ferrite and cones form a "toothpick" assembly. The input and output waveguides preserve polarization orientation. The ferrite itself is surrounded by a superconducting solenoid and the entire toothpick assembly acts as a dielectric waveguide, allowing hybrid electric modes to propagate. Details of the design and construction of FRMs can be found in \cite{Keating09}.  

Polarization rotation takes place only within the cylindrical ferrite. Bias currents driven through the solenoid generate a longitudinal magnetic field within the FRM, rotating the polarized input by an angle
\begin{equation}
\theta = VlB,
\end{equation}
where $V$  is the Verdet constant, a parameter describing the intrinsic polarization properties of the ferrite material, $l$ is the length of the ferrite cylinder, and $B$ is the strength of the applied magnetic field. Equation 1 is known as the second-order magneto-optical Faraday effect. 

The FRMs were optimized for use in BICEP, a ground-based CMB polarimeter designed to
measure the B-mode polarization of the CMB at degree angular scales. BICEP is an on-axis refracting telescope with a 0.25 m aperture capable of scanning in azimuth and elevation and rotating about the boresight. Full details of the BICEP instrument can be found in \cite{Yoon06} and \cite{Takahashi09}. Here we provide only a brief summary. 

BICEP's focal plane consists of 49 pairs of orthogonally oriented polarization sensitive bolometers (PSBs) \citep{Jones03} cooled to 0.25 K. During the 2006 observing season, 25 PSB pairs were tuned for band center of 100 GHz (beam size 0.93 degrees) and the other 24 for 150 GHz (beam size 0.60 degrees). Each pixel is individually coupled to a stack of three corrugated microwave feed horns cooled to 4K. The upper edge of the frequency pass band is defined by a series of metal mesh resonant filters placed in front of the feed horn. The lower edge is defined by the waveguide cut-off imposed by the horn itself. 

\begin{figure}
  \centering
  \includegraphics[scale=0.65]{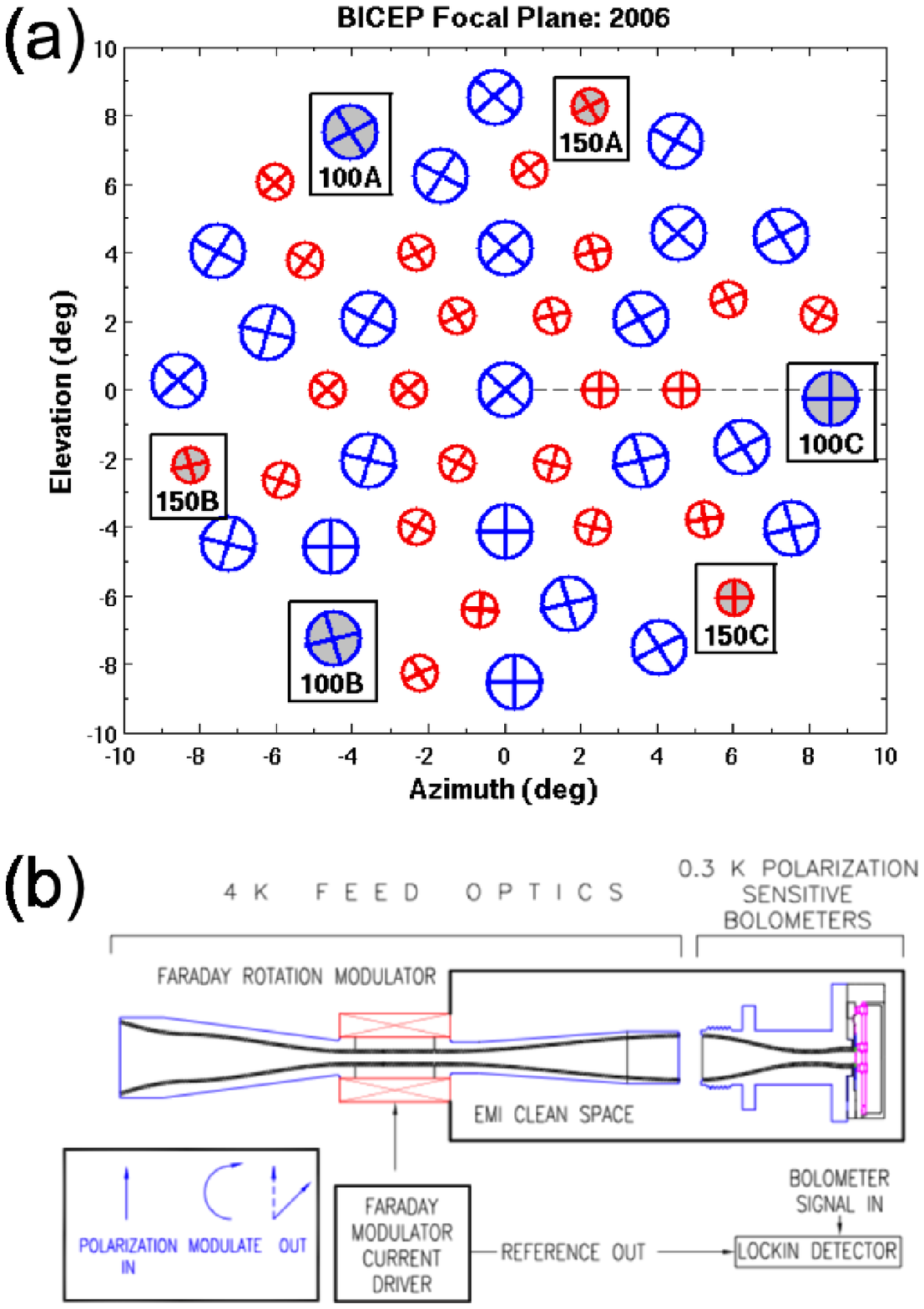}
  \caption[FRMs in BICEP]{(a) The BICEP focal plane for 2006. Each circle represents a pixel with a pair of orthogonal bolometers with polarization sensitivity axes depicted by +. The larger blue circles represent 100 GHz pairs and the smaller red circles represents 150 GHz pairs. The dashed line represents the boresight reference angle of zero. The circles that are shaded in gray in boxes around the perimeter feature FRMs. (b) The relative location of FRMs within the optical path. The FRMs are housed at 4 K between the feedhorn and bolometer enclosure. The bolometer signal is demodulated using lock-in amplification with phase referenced to the driving current.}
\label{focalplane}
\end{figure}

The FRMs were present in the optical path of six PSB pairs around the perimeter of the focal plane: three at each of 100 GHz and 150 GHz. Figure~\ref{focalplane} shows the location of the FRMs in the focal plane of BICEP along with a cross section of the FRM positioning in the optical path. For the remainder of the paper the FRMs at 100 GHz will be referred to as 100A, 100B, and 100C and at 150 GHz as 150A, 150B, and 150C as labeled in Figure~\ref{focalplane}(a). 

\section{Instrument Characterization}

The FRMs were subjected to a rigorous series of lab and field tests to characterize their behavior and choose optimal operating parameters for BICEP observations. The following section describes the FRM laboratory testing and results. 

\subsection{Rotation Angle and Bias Signal}

\begin{figure}
  \centering
  \includegraphics[scale=0.45]{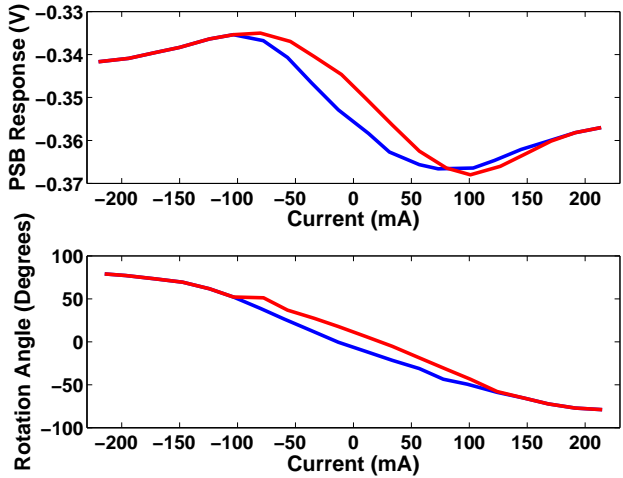}
  \caption[Rotation Angle Calibration]{Raw detector response of a PSB pair (upper two plots) and calculated rotation angle of the FRM (lower to plots) is shown for a single FRM calibration test. The red and blue curves correspond to increasing and decreasing FRM bias current respectively. FRM hysteresis can clearly be seen in both plots. The hysteresis curve separation at zero bias current was between $10^{\circ}$ and $25^{\circ}$.}
\label{rotangle}
\end{figure}

FRMs modulate mm-wave signals by rotating the direction of polarized radiation relative to the axis of polarization of the corresponding bolometers. To mitigate systematic offsets, this relative rotation must be well-calibrated and time-invariant. Measurements of the rotation angle as a function of bias current were made by placing an aperture-filling wire grid polarizer (resulting in a 100\,\% linearly polarized signal) at the telescope window, biasing the FRM with a triangle wave signal, and measuring the voltage response of the PSB pair beneath the rotator. An example output of this technique is given in Figure~\ref{rotangle}. 

As the FRM bias current is increased and decreased, the bolometer voltage responds with a sinusoidal wave form that ``flattens" at the turnarounds when the ferrite begins to saturate magnetically. From this data, the relative rotation angle of the FRM, $\theta$, can be calculated by
\begin{equation}
\theta = \frac{1}{2}\sin^{-1}\Big{(} \frac{V - (V_{\max} + V_{\min})/2 }{(V_{\max} - V_{\min})/2} \Big{)}
\end{equation}
where $V$ is the bolometer response voltage and $V_{\max}$ and $V_{\min}$ are the maximum and minimum of that voltage, respectively. Figure~\ref{rotangle} shows the voltage response curve as a function of $\theta$. FRMs in BICEP can continuously rotate a polarized signal through a range of approximately $\pm 80^{\circ}$. Maximum rotation corresponds to a bias current of approximately 300 mA.

The curves shown in red in Figure~\ref{rotangle} are for increasing bias whereas the curves shown in blue are for decreasing bias. The two curves do not overlap due to magnetic hysteresis, where a change in magnetization will lag with respect to the externally applied field due to configurational internal forces. Magnetic hysteresis prevents assigning a one-to-one correspondence between current and rotation angle, creating a degeneracy in the bias angle for values of the bias current where hysteresis is present. Additionally, the degree of hysteresis was shown also to be a function of bias frequency, further increasing the complexity of demodulation. To avoid the complications of magnetic hysteresis the FRMs were biased with a square wave signal. This limited the FRM rotation angle to only two states, which could be uniquely determined by the magnitude and sign of the bias current. 

The amplitude of the bias signal was chosen such that the FRMs would have angular modulation of 45 degrees, corresponding to a bias current of approximately $\pm$125 mA. In the reference frame of the bolometer, this is equivalent to switching between Stokes +Q and -Q, effectively transforming \emph{a single} PSB into two orthogonal detectors. The square wave bias was tuned for a frequency of 1 Hz, slightly higher than the 1/$f$ knee for the BICEP temperature data, generally 0.5-1 Hz for a single pixel \citep{Takahashi09}. During unmodulated CMB and Galactic half-scans, BICEP shifts the $1/f$ knee by scanning the telescope at a scan speed of $2.8^{\circ}/s$. Although the FRMs are capable of switching at frequencies orders of magnitude higher, the FRMs in BICEP were limited by the bolometer time constants, which had a median of 21 ms across the focal plane. A 1 Hz bias was found to be the best compromise between maximizing integration time per cycle, which lowers the frequency but decreases the noise in demodulation (Section 4.2.3), and minimizing the $1/f$ atmospheric fluctuations. 

\subsection{Instrumental Polarization}

Instrumental polarization (IP) is spurious polarization detected when only unpolarized light is observed. IP can result from use of the FRMs if FRM toothpick assembly is tilted relative to the optical axis of light to the detector. The FRM's IP were measured by placing unpolarized aperture-filling 300 K and 77 K sources at the cryostat window and biasing the FRMs through their full range of rotation angles. The resultant fractional IP is defined as
\begin{equation}
\mathrm{IP} = \frac{1}{2} \Big{(} \frac{V_{AC}(300K)}{V_{DC}(300K) - V_{DC}(77K)} \Big{)} \Big{(} \frac{300K - 77K}{300K} \Big{)}
\end{equation}
where $V_{DC}(300K)$ is the mean bolometer voltage at 300 K, $V_{DC}(77K)$ is the mean at 77 K, and $V_{AC}(300K)$ is the peak-to-peak bolometer voltage at 300 K.

Individual FRM pixels were found to have a repeatable and time-invariant IP signal, with a standard deviation of approximately six percent (and limited to ten percent) over the testing period of several months. The scatter of the mean IP between FRM assemblies, however, varied considerably. The average value of all FRM assemblies was found to be 0.59\% with a standard deviation of 0.43\%. The maximum value among all the FRM pixels was found to be less than 1.2\%, for FRM150A. 

\subsection{Rotation Angle Calibration}

Two rotation angle calibration runs were performed with the BICEP FRMs. A dielectric sheet calibrator (DSC) was used based on the design from POLAR \citep{O'Dell02, Takahashi08}. The DSC consisted of an eccosorb-lined cylinder with a polypropylene sheet mounted at $45^{\circ}$ that served as a beam splitter, creating a small polarized signal from input ambient temperature and sky loads. The direction of the polarized signal was determined by the angle of the DSC relative to the detector. The DSC was placed above BICEP's optical window, and as the boresight was rotated each PSB exhibited a sinusoid-like response that varied as a function of radial position in the focal plane. Model fits using the known properties of the DSC yielded detector polarization angles for each PSB in the focal plane. 

\begin{figure}
 \centering
  \includegraphics[scale=0.43]{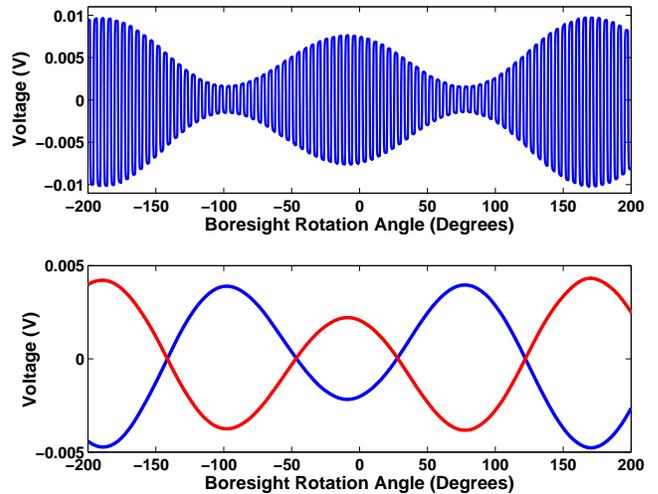}
  \caption{Dielectric sheet calibration for FRM150A. The top panel shows the 1 Hz FRM modulated signal. The bottom panel shows the split demodulated timestreams at $\pm45$ degrees.}
\label{yukical}
\end{figure}

Figure~\ref{yukical} shows the timestream response of a BICEP PSB to DSC calibration with FRM modulation. To extract polarization angle, the FRM output was demodulated into two separate timestreams as described in Section 4.2.3. The result is shown in the bottom panel of Figure 4. Each timestream was fit to the DSC model for detector angle. Given a nominal PSB orientation angle $\psi$, derived from DSC tests in the absence of modulation, the two demodulated timestreams with proper bias should yield detector angles $\psi \pm 45^{\circ}$. 

DSC calibration showed proper bias rotation for four of the six FRMs. One of the rotator biases was incapable of supplying the currents necessary for full $\pm 45^{\circ}$ rotation (FRM 100B) and another was deemed unreliable based on the high standard deviation in its rotation angle (FRM 150C). In addition, the PSB pair associated with FRM 100C had excess noise in both the modulated and demodulated timestreams, rendering it ineffective for celestial polarization studies.

The results of the DSC calibration are summarized in Table~\ref{yukitable} for all six PSB pairs. The standard deviations listed in the table reflect a true deviation of the rotation range during bias periods and are not limited by measurement readout accuracy. 

\begin{deluxetable}{cccccccc} 
\tablecolumns{3} 
\tablecaption{DSC calibration results for all six PSB pairs. The table lists the total rotation modulation of the polarization angle. For perfect $\pm 45^{\deg}$ polarization modulation, 90$^{\circ}$ is expected. Polarization modulation in the plus and minus direction is one-half of the listed value. The standard deviation is calculated across each individual run and averaged over the two different rotation angle calibrations.} 
\tablehead{
\colhead{FRM PSB} & \colhead{Avg Rotation Range $\pm$ Standard Deviation ($^{\circ}$)}}
\startdata 
100A & 85.5 $\pm$ 7.3 \\ 
100B & 41.4 $\pm$ 12.5 \\ 
100C & 79.7 $\pm$  7.4 \\
150A & 90.2 $\pm$  0.1 \\ 
150B & 88.2 $\pm$  4.5 \\ 
150C & 73.8 $\pm$ 18.8  
\enddata 
\label{yukitable}
\end{deluxetable} 

\section{Galactic Observations}

\begin{deluxetable*}{cccccccc} 
\tablecolumns{8} 
%\tablewidth{0pc} 
\tablecaption{Details of each FRM Observation, ordered sequentially. Each observation was segmented into 9-hr sections described by the No. Sections column. For each of these sections, the rotation angle of the boresight is described. The next two columns describe the range of elevation and step size at the boresight center. Azimuth parameters include the length of all azimuthal half-scans, the scan speed, and the number of uni-direcational half-scans performed at each step in elevation.} 
\tablehead{
\colhead{Date(s)} & \colhead{No. Sections}  & \colhead{Boresight Angle ($^{\circ}$)}  & \colhead{Elevation ($^{\circ}$)}  & \colhead{El Step}   & \colhead{Az Range ($^{\circ}$)} & \colhead{Scan Speed ($^{\circ}/s$)}   & \colhead{Half-Scans}}
\startdata 
April 21, 2006 & 5 & \{315, 315, 180, 180, 0\} & 55.5--60.5 &  0.25  & $\approx$77 & 0.25 & 4 \\ 
Oct. 14, 2006 & 4 & \{154, 164, -152, -142\} & 55.5--61.2 &  0.1  & $\approx$12 & 0.25 & 22 \\ 
Oct. 18, 2006 & 4 & \{154, 164, -152, -142\} & 55.5--61.2 &  0.1  & $\approx$12 & 0.25 & 22 \\ 
Oct. 28, 2006 & 4 & \{120, -150, -150, 120\} & 55.5--61.2 &  0.1  & $\approx$12 & 0.25 & 22 \\ 
Oct. 30, 2006 & 4 & \{120, -150, -150, 120\} & 55.5--61.2 &  0.1  & $\approx$12 & 0.25 & 22 \\ 
\enddata 
\label{observationtable}
\end{deluxetable*} 

\subsection{Observing Strategy}

In 2006, five observations of the Galactic plane were made with the FRMs biased. The target fields are shown in Figure 5. The first of these observations, referred to as the ``shallow" observation, is defined by the celestial coordinate system $(\alpha, \delta)$ = (Right Ascension, Declination) as $180^{\circ} < \alpha < 290^{\circ}$ and $-70^{\circ} <  \delta <-45^{\circ}$ (Figure 5). The shallow FRM observation was used mainly for diagnostic purposes; this was the first test of FRM modulation on a celestial source. 

All deep observations focused on a small portion of the Galactic plane $238^{\circ} < \alpha < 248^{\circ}$ and $-53^{\circ} < \delta < -50^{\circ}$. The small area was necessary to maximize integration time, as the goal of the deep observations was a detection of polarization from a celestial source. Due to their respective positions in the BICEP focal plane, no two FRMs could scan this same region of sky simultaneously, so integration time was split evenly between two FRMs: FRM100A and FRM150A. FRM100A was chosen because it had the highest optical efficiency, lowest IP, and a reliable current bias signal. FRM150A was chosen for its extremely consistent bias signal during DSC calibration and its close proximity to FRM100A in the focal plane. An integration time of approximately 72 hours on this deep patch was achieved for each FRM used for analysis. 

For all FRM observations, azimuth-elevation raster scans were used. At each step in elevation, the telescope performed a number of unidirectional scans in azimuth (known as ``half-scans") back and forth across the target patch with a scan speed $0.25^{\circ}/s$. The elevation step size and number of half-scans at each step were altered between the shallow and deep observations. These values are summarized in Table~\ref{observationtable} along with the other pertinent features of each observation. During the deep observations, a single boresight rotation angle corresponds to a single FRM pixel focused on the Galactic region of interest. 

\begin{figure}
\centering
\includegraphics[scale=0.35]{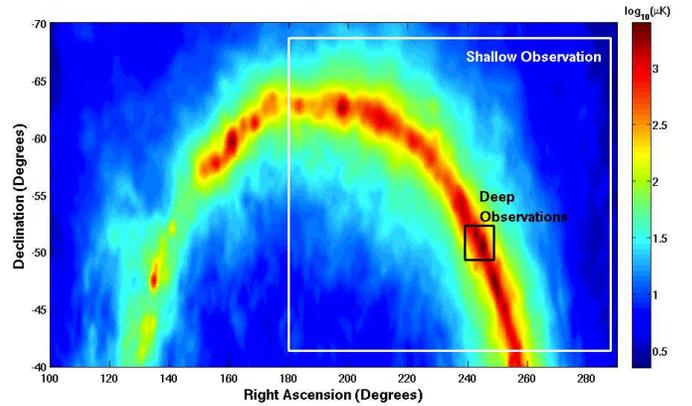}
\caption{100 GHz FDS Model 8 Galactic dust emission prediction \citep{Finkbeiner99} plotted in celestial coordinates on a logarithmic scale. The outlined regions show the areas of integration for the shallow and deep FRM observations.}
\end{figure}

\subsection{Timestream Processing}

When characterized by Stokes T, Q, and U parameters, the time domain response of an unmodulated BICEP PSB as a function of frequency $\nu$ and direction $\Omega$ is modeled by
\begin{eqnarray}
d(t) = K_t \otimes \Big{\{} n(t) + g \int d\nu A_eF(\nu) \int d\Omega P(\Omega) \nonumber\\ \Big{[} T + \frac{1-\epsilon}{1+\epsilon}\Big{(}Q(\cos (2\psi) + U\sin (2\psi) \Big{)}\Big{]}\Big{\}}, 
\end{eqnarray}
where $\psi$ is the polarization orientation angle of the PSB, the parameter $\epsilon$ quantifies the amount of cross polar leakage of the PSB, $P(\Omega)$ is the beam response function, $F(\nu)$ is the spectral response, $A_e$ is the effective antenna area, $g$ is the gain factor that converts voltage to temperature, $n(t)$ is a term describing total atmospheric and detector noise, and $K_t$ is the time domain impulse response of the detector. A complete description of the measurements and experimental procedures used to obtain these terms is given in \cite{Takahashi09}. 

\subsubsection{Preliminary Processing}

Data processing begins by deconvolving the complete, measured transfer functions for all FRM timestreams. The beginning and end of each half-scan is identified and periods of non-zero telescope mount acceleration are trimmed from the end points. Horizon and celestial boresight coordinates are calculated by applying a pointing model to the raw boresight pointing using the focal plane coordinates of the pixel of interest.

\subsubsection{Relative Gains}

Relative detector gains are applied by fitting to the elevation nods, a period at the beginning of each fixed elevation scan when the telescope performs a rounded triangle wave motion with an elevation amplitude of 1.2 degrees and a duration of approximately 27 seconds. During this time, for consistency, the FRMs are modulating the sky signal. The responsivity factor for each PSB is found by fitting the detector voltage versus the changing line of atmospheric sight, given by the cosecant of elevation multiplied by the scale height. Before applying the relative gains, each PSB gain factor is weighted by the average of the PSB pair during the scan set. 

\subsubsection{Demodulation}

Although many demodulation techniques were considered, time domain demodulation was used  due to its simple and robust properties and accurate polarization reconstruction. Figure 6 is a graphical example of the timestream demodulation using FRM 150A. 

The first step of demodulation was to clean the bias current signals of software spikes and glitches via nearest-neighbor interpolation. The current signal then displayed fluctuations on the $<1\%$ level, and so perfect $\pm45^{\circ}$ rotation was assumed for positive and negative values of the bias, respectively. 

\begin{figure}
\centering
\includegraphics[scale=0.46]{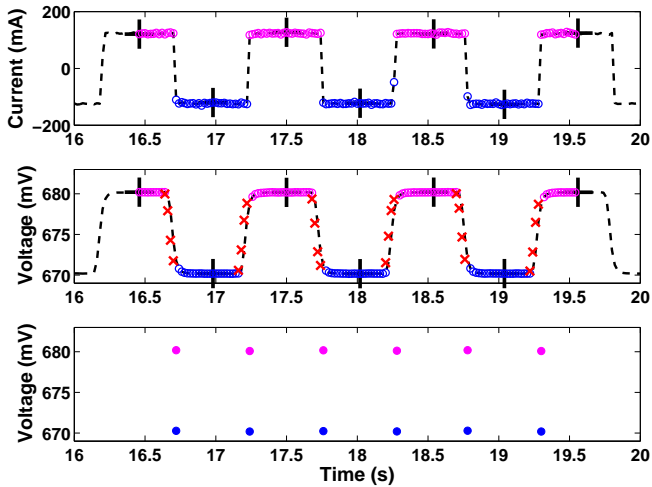}
\caption{Illustration of time domain demodulation. The top plot shows the bias current as a function of time. Points in pink are positively biased, corresponding to a $+45^{\circ}$ rotation. Points in blue are negatively biased, corresponding to $-45^{\circ}$ rotation. The black crosses are the midpoints of each bias plateau. The middle plot shows the corresponding points mapped onto the bolometer timestream. The red crosses are points of gradual switching resultant from bolometer time constants, and are cut from analysis. The remaining points are averaged together for each half-plateau and are stored as two new demodulated timestreams as shown in the lowest panel. The pink filled circles comprise the demodulated $+45^{\circ}$ timestream, whereas the filled blue circles correspond to $-45^{\circ}$ timestreams.}
\end{figure}

The current biases were then used as the lock-in phase reference signals for demodulation of the corresponding bolometer timestreams. All points at which the current bias transitioned across zero were identified and the median number of points between transitions was computed. Any region where the switching period differed from the median by more than 10\%, along with one plateau region on either side, was identified and the corresponding region in the bolometer timestream was excluded from analysis. This resulted in a loss of $<1\%$ of the usable timestream data as the bias signal exhibited very few deviations from normal periodicity. 

The midway point between each two adjacent transitions in the bias signal was identified, dividing each ``bias plateau" into two halves. The cleaned bias signal and its segmentation are shown in the top panel of Figure 6. The average value of the corresponding bolometer timestream, along with other pertinent variables (pointing, time, etc.), for each half was computed, less the two to four points closest to the transition to account for the non-instantaneous bolometer response. Although deconvolution using the measured time constant does recover a small number of points along each transition -- without deconvolution, a minimum of six points were cut on each side -- the bolometer response to bias switching is still imperfect and a fraction of the data surrounding each transition must still be removed from analysis. The number of points cut from each side of the transition was derived independently for each bolometer using iterative fits to the elevation nods. To maximize the signal-to-noise, the minimum number of data points was cut that yielded a convergent fit to the elevation nod signal. 

The bolometer timestream averaged values were then separated based on the sign of the corresponding bias signal. Essentially, the timestream of a single bolometer at polarization angle $\psi_{ref}$ has been demodulated and downsampled into two demodulated bolometer timestreams at $\psi_{ref}+45^{\circ}$ and $\psi_{ref}-45^{\circ}$. This is shown in the last panel of Figure 6. The sum and difference of the two demodulated timestreams from each individual PSB was then calculated. 

\subsubsection{Masking and Filtering}

\begin{figure}
\centering
\includegraphics[scale=0.54]{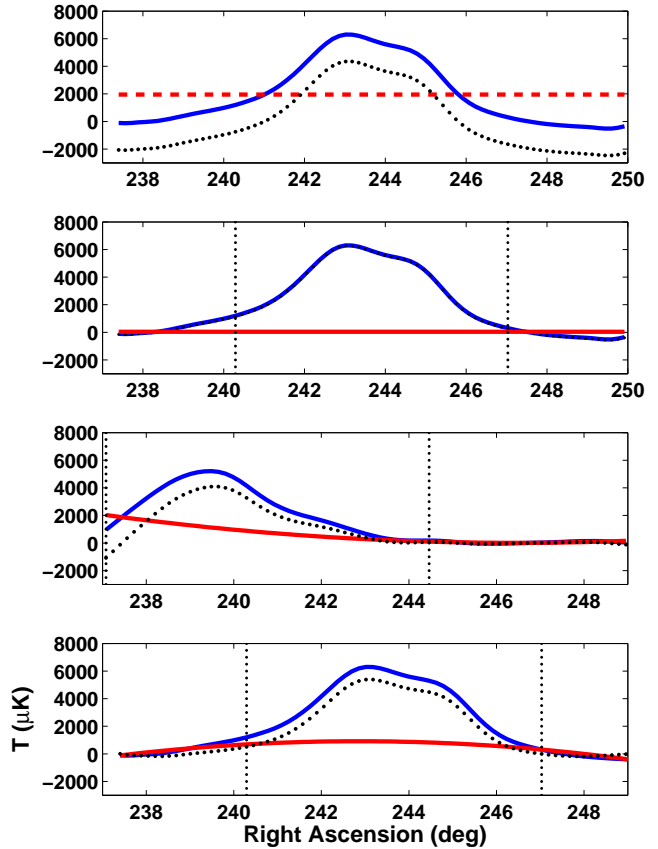}
\caption{Summed detector timestreams for FRM deep integration Galactic half-scans with different filtering methods. The top plot shows the timestream (blue), DC offset fit (red), and the resultant filtered timestream (black). Note the severe distortion from fitting the Galactic region. The second plot shows the same data with Galactic intensity removed from the fit. The boundaries of the mask are shown by black dotted vertical lines. The DC offset is subtracted from the masked region, leading to minimal artifacts. The third plot shows data that terminates on the masked Galactic region, fit with a higher-order polynomial; the interpolation diverges through the Galactic plane. The final plot illustrates the effects of subtracting a higher order polynomial fit from data with an insufficient mask to completely remove Galactic signal.}
\label{filtering}
\end{figure}

A DC-offset was removed from both the sum and difference timestreams via mean-subtraction of each azimuthal half-scan. An obvious distortion arises when fits for the mean include the strong  Galactic signal, so the Galaxy was masked from the fits. The top two panels of Figure~\ref{filtering} show an example of such distortion and the lack of distortion in the masked fit.

BICEP CMB half-scans are filtered using a third order polynomial \citep{Chiang2010} and BICEP Galactic half-scans utilize a second order polynomial \citep{Bierman11} in order to accurately remove atmospheric $1/f$ noise. For the deep FRM observations, the length of the half-scans across the galaxy prohibited such a filtering scheme. A mask large enough to remove all Galactic signal from the fits resulted in the majority of half-scans beginning or terminating on the masked region. Interpolated fits that are not constrained by data on both sides of the mask tend to diverge and no longer reasonably approximate noise within the mask. Conversely, smaller masks avoid this problem but leave residual Galactic signal for the polynomial fit, causing higher order polynomial fits to remove true Galactic signal instead of only noise. These results are shown in the last two panels of Figure~\ref{filtering}. While mean subtraction does not remove large-scale noise from the data, FRM modulation mitigates some of the $1/f$ atmospheric fluctuations that would otherwise need to be filtered. 

The optimal mask cuts maximal Galactic signal from the fits while retaining sufficient off-Galactic data to yield statistically significant fits for the mean. Given the Galactic coordinates defined by $(b, l)$ =  (latitude, longitude), the radius of the masked region was reduced from $\mid b \mid < 3^{\circ}$ (the maximal possible mask for the deep integration scans) until the DC offset fits converged for the majority of half-scans. The optimal mask was found to be $\mid b \mid < 1.5^{\circ}$.

Following DC offset subtraction, residual high-frequency power from demodulation was removed via a Butterworth low pass filter at 0.6 Hz. A data quality check was then performed where any half-scan with a signal spike more than 7 times the standard deviation was removed from analysis. These regions comprised $<1\%$ of the total usable data. 

\subsubsection{Instrumental Polarization Removal}

Temperature to polarization leakage from a bright source such as the Galaxy can cause significant artificial structure in polarization. Laboratory testing revealed that FRM150A showed approximately 1\% IP, the highest of any of the devices; a value that could distort Galactic polarization maps. To remove this temperature to polarization leakage, the IP value for each half-scan was calculated using a fit to the preceding elevation nod. The derived fractional IP value was then multiplied by the half-scan sum data and subtracted from the difference data. 

\begin{figure}
 \centering
  \includegraphics[scale=0.5]{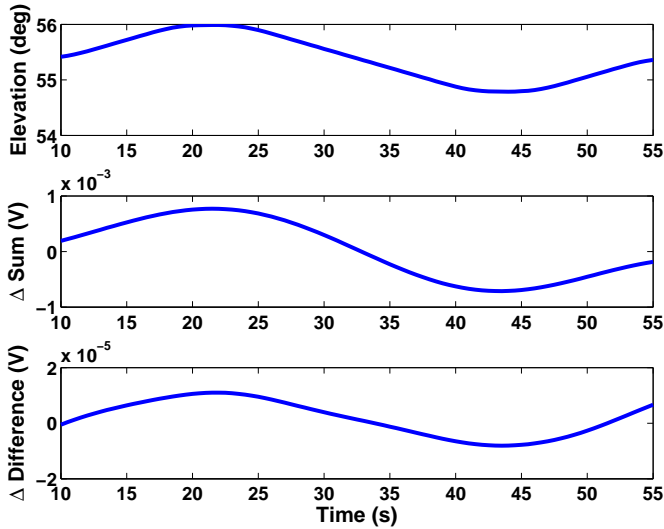}
  \caption{From top to bottom: Elevation angle, FRM150A demodulated sum, and FRM150A demodulated difference data for an elevation nod versus time for FRM 150A. The mean has been subtracted from both the sum and the difference data so that small changes can be seen. The difference data is approximately 1\%, of the sum data, indicative of instrumental polarization.}
\label{IPplot}
\end{figure}

Figure~\ref{IPplot} shows an example of IP leakage derived from elevation nods. Because the atmosphere is unpolarized, the elevation nods should cause a change in the summed (but not differenced) data for each demodulated PSB. The data from each elevation nod for each PSB was demodulated, summed/differenced, mean subtracted, and low-pass filtered. The summed and differenced timestreams are both fit to the cosecant of elevation to derive average responses to the changing line of sight. Average IP values were computed by taking the mean of the absolute value of the difference divided by the sum across all scans and all runs. The average values at both frequencies across all observations was found to be IP = 0.21\,\% and IP = 0.88\,\% at 100 GHz and 150 GHz respectively.

To confirm that the elevation nods revealed IP and not some other source of spurious polarization, the fit values for both the sum and difference were plotted as a function of elevation. The summed data increased with elevation due to greater atmospheric loading on the PSBs, and the differenced data was shown to scale proportionally to the sum. The correlation between summed and differenced data indicate true IP. 

\subsubsection{Map Making}

After timestream data processing was complete, Stokes $T$, $Q$, and $U$ were derived following the formalism in ~\cite{Jones07}. Temperature data were obtained by binning the filtered detector timestreams into map pixels, $\bf{p}$, as 

\begin{displaymath}
T(\bf{p}) = \Big{(} \sum\limits_{i}^n \sum\limits_{j \epsilon \bf{p}} w_{ij}^{+}d_{ij}^{+} \Big{)} \Big{/} \Big{(} \sum\limits_{i}^n w_{ij}^{+} \Big{)},
\end{displaymath}
where the indices $i$ and $j$ denote the PSB channel number and timestream sample number respectively. The variable $d^+$ is the sum demodulated timestream, $n$ is the number of PSBs, and $w^+$ is the weight assigned to each demodulated sum. Stokes $Q$ and $U$ were calculated by
\begin{eqnarray}
\sum\limits_{i}^n \sum\limits_{\j \epsilon \bf{p}} w_{ij}^{-}
\left( \begin{array}{c}
d_{ij}^{-}\alpha_{ij} \\
d_{ij}^{-}\beta_{ij} \end{array} \right) = \nonumber\\ \frac{1}{2}
\sum\limits_{i}^n \sum\limits_{j \epsilon \bf{p}} w_{ij}^{-}
\left( \begin{array}{cc}
\alpha_{ij}^2 & \alpha_{ij}\beta_{ij} \\
\alpha_{ij}\beta_{ij} & \beta_{ij}^2 \end{array} \right)
\left( \begin{array}{c}
Q(\bf{p}) \\
U(\bf{p}) \end{array} \right) 
\end{eqnarray}
where $w^-$ is the weight assigned to the demodulated timestream difference and $\alpha$ and $\beta$ are the differenced orientation factors
\begin{eqnarray}
\alpha_{ij} = \gamma_{i} (\cos[2(\psi_{ij} + 45^{\circ})] - \cos[2(\psi_{ij} - 45^{\circ})])  \\
\beta_{ij} = \gamma_{i} (\sin[2(\psi_{ij} + 45^{\circ})] - \sin[2(\psi_{ij} - 45^{\circ})]) 
\end{eqnarray}
The factors of $\pm 45$ degrees come from the Faraday rotation of the nominal polarization angle of the $i^{th}$ PSB. The same correction factor describing cross polar leakage, $\gamma_{i} = \frac{1-\epsilon_i}{1+\epsilon_i}$, is applied to both terms $\alpha_{ij}$ and $\beta_{ij}$.  Equation 5 is degenerate for a single FRM PSB at time $j$. To break the degeneracy, the same sky pixel $p$ is observed by the same modulated detector at a different polarization angle $\psi_{ij}$. With more than one observing angle the off-diagonal elements of Equation 5, $\alpha_{ij}\beta_{ij}$, average to zero and the matrix can be inverted to solve for $Q$ and $U$. 

For each half-scan the inverse of the variance of the pair sum and difference timestreams is used as the weight $w^{\pm}$. Total integration time is also computed for each pixel. 

\subsubsection{Absolute Calibration}

To relate detector units to CMB units, BICEP measurements are cross-correlated with WMAP temperature maps to derive an absolute gain calibration. A complete description is given in \citep{Chiang2010}. 

This same comparison cannot be made for the FRMs, as the FRMs were never biased and observing during CMB observations. Instead, FRM detectors were calibrated via comparison with the calibrated 3-year BICEP temperature maps of the bright arm of the Galaxy. The absolute calibration factor was computed as the slope of a linear fit to the pixel-pixel scatter plot of the two maps. The variance of the difference between the fit data and the BICEP data is used to compute the uncertainty. This method yields values of $1.1288 \pm 0.0023$ and $1.0011 \pm 0.0017$ times the nominal absolute calibration factors for BICEP at 100 and 150 GHz respectively. 

\section{Map Results}

\begin{figure}
\centering
\includegraphics[scale=0.32]{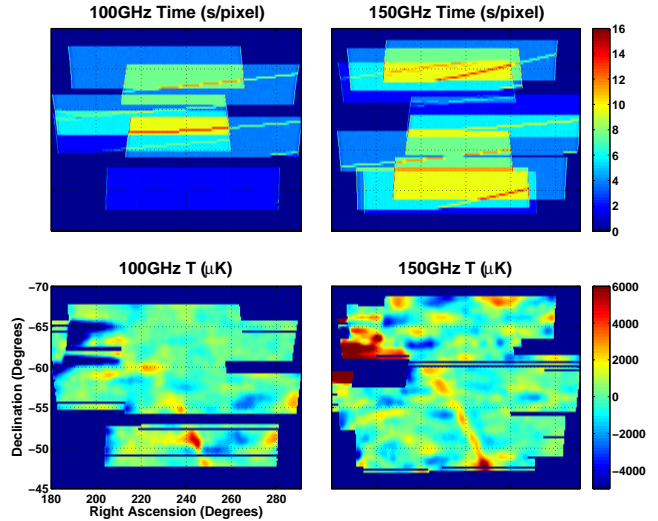}
\caption{Integration time and temperature maps from the shallow FRM observation on April 21, 2006. The temperature of the Galactic plane can clearly be seen above the residual noise. The large scale noise features at right ascension $180^{\circ}-200^{\circ}$ and declination of $\approx-63^{\circ}$ are an artifact of masking and filtering (Section 4.2.4). Although the non-FRM PSBs are observing during this shallow observation, the slow scan speed places their observing band well below the 1/$f$ knee, leading to excessive noise. For this reason, these maps are intentionally omitted from further analysis.}
\label{ShallowInt}
\end{figure}

Figure~\ref{ShallowInt} shows both the integration time and Galactic temperature maps derived from the shallow FRM observation at 100 GHz and 150 GHz. Because the shallow observation featured long sweeps of the Galaxy, half-scans were filtered using a second order polynomial rather than mean subtraction. The maps were binned with Healpix \citep{Gorski05} pixelization $0.25^{\circ}$. A $\sigma = 0.5^{\circ}$ Gaussian smoothing function has been applied to all maps. 

Due to glitches in the rotator bias signals, only the first and last 9 hour sections resulted in usable data, a total of approximately seventeen hours of integration time including data cuts. The deepest integration time was $\approx 15 s$ per pixel, with an average of 4.1 and 5.4 $s$ per pixel at 100 and 150 GHz, respectively.  Data from 5 out of 6 of the FRM pixels (two at 100 and three at 150) were used for these maps; one was omitted due to biasing problems.  

The temperature maps from this observation reveal bright Galactic emission confined mainly to $\mid b \mid<1^{\circ}$, with both large scale features and compact sources. The intensity is  greater at 150 GHz than at 100 GHz, as predicted by models of dust in the interstellar medium \citep{Finkbeiner99}, and consistent with other recent mm-wave experiments \citep{Culverhouse10,Bierman11}. Although the off-Galactic portions of the map show relatively large noise fluctuations due to the low integration time, the Galactic signal dominates the noise for both bands. This is, however, not true for the much fainter polarization signal (not pictured); a detection of statistically-significant polarization was not seen in this observation. The large scale noise features that differ between the 100 GHz and 150 GHz maps result from a combination of $1/f$ atmospheric noise and beam smoothing. 

\begin{figure*}
\centering
\includegraphics[angle=90,scale=0.28]{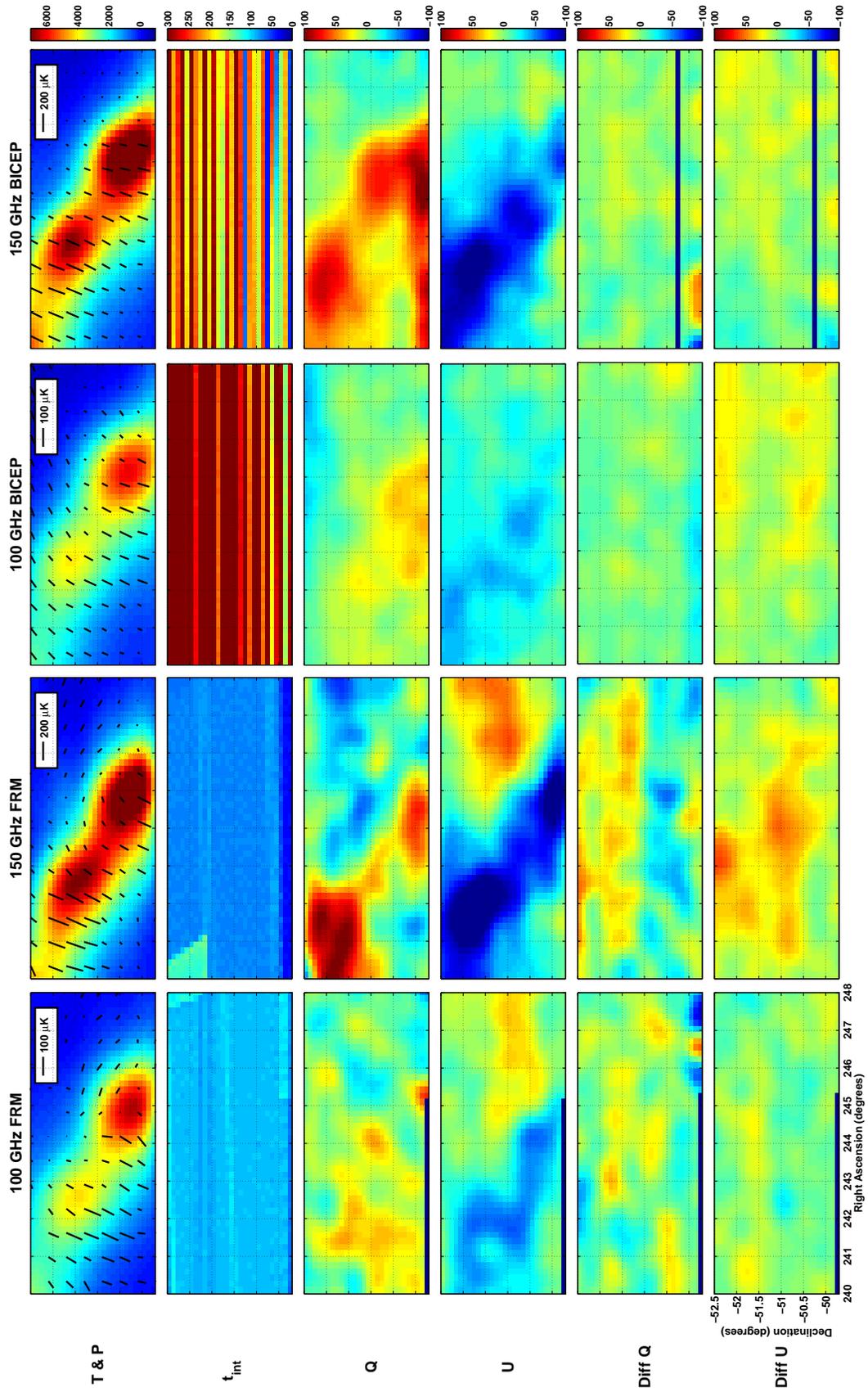}
\caption{FRM and BICEP maps of the deep integration region. From top to bottom, each column displays a single plot of temperature with polarization vectors (T \& P), integration time (T$_{\mathrm{int}}$), Stokes Q, Stokes U, and Q and U scan direction difference maps (Diff Q and Diff U). All maps are displayed in $\mu$K except for integration time, which is displayed in seconds. The first two columns are the FRM deep observation results at 100 and 150 GHz. The last two columns show maps made with the nominal BICEP observations using non-FRM PSBs at 100 and 150 GHz.}
\label{deepint}
\end{figure*}

The left two columns of Figure 10 show integration time as well as the Galactic T, Q, U, polarization vector, and Q and U difference maps at both 100 GHz and 150 GHz from the FRM deep observations. Only two FRMs, 100A and 150A, were used to accumulate the maps. Maps are binned into Healpix \citep{Gorski05} $0.1^{\circ}$ pixels and high-pass filtered via DC offset subtraction (section 4.1.1). A $\sigma = 0.2^{\circ}$ Gaussian smoothing kernel has been applied to the maps at both frequencies. 

The integration time in this region averages 77.7 and 67.2 seconds per map pixel at 100 and 150 GHz respectively. The T, Q, U maps all show an increase in signal from 100 to 150 GHz. At both frequencies, large-scale positive Q and negative U polarization can be seen along the plane of the Galaxy. These polarization features are generally confined to within $\mid b \mid <1.5^{\circ}$ of the center of the Galactic plane. The polarized portions of the maps are coincident with the largest magnitude intensity in Declination and slightly offset towards lower Right Ascension. At both frequencies, the magnitude of the signal in Stokes U is approximately five times larger than the signal in Stokes Q. The average signal in Stokes U confined to $\mid b\mid < 1.5^{\circ}$ is given by -16.2 $\mu K$ and -48.4 $\mu K$ at 100 and 150 GHz respectively. Given the error bars defined by difference maps (Section 5.2), this corresponds to an average S/N ratio of 1 and 3 in each smoothed map pixel for the average galactic polarization signal seen at 100 and 150 GHz respectively.

BICEP observations of the same region were made during the 2006 - 2008 BICEP observing seasons and compared to the FRM data. The BICEP Galactic scan strategy uses long sweeping scans across the entire Galactic plane at a step size of $0.25^{\circ}$, prohibiting a direct comparison of a single non-FRM PSB pair to a single FRM PSB pair; for a single non-FRM PSB pair, the integration time is too low and the number of boresight angles is insufficient to accumulate polarization. As such, maps were accumulated using all non-flagged PSB pairs in the focal plane. PSB pairs containing FRMs (among others) were flagged and excluded from analysis. The data were taken over all three BICEP observing seasons, totaling 763 observing hours for the entire shallow observation region.  Non-FRM data were processed using the same analysis pipeline as the FRM data with a few exceptions. The initial deconvolution was applied with a low-pass filter at 5 Hz. Following deconvolution, the data were downsampled to 10 Hz before preliminary processing and relative gain calibration. The demodulation step was omitted; the sum and difference data were taken between the two PSBs within each pair. The data were then subjected to the same masking and filtering strategy as the FRM pixels, with the additional complication that all data outside the region of deep integration were also masked from the fits. This is to keep large portions of off-Galactic data from weighting the DC offset subtraction, which was not possible with the FRM deep integration scan strategy. Maps were accumulated using the same formalism as Section 4.2.6, except Equations (6) and (7) are modified to
\begin{eqnarray}
\alpha_{i,j} = \gamma_{i} (\cos[2(\psi_{i+1,j})] - \cos[2(\psi_{ij})])  \\
\beta_{ij} = \gamma_{i} (\sin[2(\psi_{i+1,j})] - \sin[2(\psi_{ij})]) 
\end{eqnarray}
 where $i$ and $i+1$ are the indices of the two PSBs within the pair that have been subtracted. The binning and smoothing of the maps is identical to the FRM deep observations. 

The right two columns of Figure 10 show the integration time as well as Galactic T, Q, U,  polarization vector, and Q and U difference maps at both 100 GHz and 150 GHz from the BICEP observations accumulated using the non-flagged focal plane PSBs. A total integration time of approximately 324 hours and 194 hours was taken at 100 and 150 GHz respectively, with a mean integration time of 412 seconds per pixel at 100 GHz and 246 seconds per pixel at 150 GHz. The maps exhibit very good agreement with the FRM maps in both temperature and polarization at both frequencies. As evidenced by the polarization vector maps in Figure~\ref{deepint}, both data sets show the same strong Galactic polarization, roughly perpendicular to the Galactic plane, with the highest polarized signal in the upper left quadrant. Both the FRM and non-FRM BICEP maps exhibit a diminution of polarization across the Galaxy with increasing RA. At each frequency and for all polarization maps, the contribution of Stokes U dominates the polarized signal. Both the FRM and non-FRM BICEP maps are consistent with the polarized emission expectations based on Galactic plane maps published previously by BICEP and other recent mm-wave experiments \citep{Hildebrand99, Mejia05, Miville08, Culverhouse10, Bierman11}.

The polarization fraction is found by calculating the median value of the quotient of polarization and temperature across all map pixels where $\mid b \mid <1.5^{\circ}$. The variance in this same region is used to compute the uncertainty on these values. The FRM maps yield a polarization fraction of $1.32\% \pm 2.17 \%$ and $2.36\%\pm0.21\%$ at 100 and 150 GHz respectively. The results derived from BICEP's non-FRM maps show strong agreement: $1.53\% \pm 0.61\%$ at 100 GHz and $2.31\%\pm 0.02\%$ at 150 GHz. For both the FRM and non-FRM BICEP maps, 150 GHz displays little variance in polarization fraction across the galactic plane, whereas the 100 GHz variance is much higher. 

\subsection{Noise and Systematics}

The deep FRM observations were analyzed for noise and other transient issues that may have resulted from the use of the devices. The results of the FRM observations were also subjected to many self-consistency checks to in order to verify the accuracy of the data presented here. Due to the low signal-to-noise achieved for the shallow observations, only the deep observations are subjected to the difference map and noise analysis.

\subsection{Difference Maps}

Difference maps, in which the map data is split in half and differenced, were used to check the self-consistency of the Galactic maps. Although some residual signal may remain due to timestream filtering effects, the expected signal of the maps is nearly zero and all large scale Galactic temperature and polarization signals should vanish. Statistical polarization errors are quantified by taking the standard deviation for all Q and U map pixels at both 100 and 150 GHz. These results are summarized in Table~\ref{diffmaptable}. For consistency with the map results, a $\sigma = 0.2^{\circ}$ Gaussian smoothing function is applied to all maps. As such, the rms map noise for each pixel in this Table~\ref{diffmaptable} is quoted for an effective pixel size of ~0.24 deg$^2$. 

\begin{deluxetable*}{ccccc} 
\tablecolumns{3} 
\tablecaption{The rms map noise for each of the three difference maps analyzed for the deep FRM observations. The rms is quoted for an effective smoothed pixel size of ~0.24 deg$^2$. For comparison, the Galactic U signal for the FRM maps was found to be -16.2 $\mu K$ and -48.4 $\mu K$ at 100 and 150 GHz respectively.} 
\tablehead{
\colhead{Difference Map Type} & \colhead{100 GHz rms ($\mu$K/pixel)}  & \colhead{150 GHz rms ($\mu$K/pixel)}}
\startdata 
Scan Direction & 12.3 & 20.5  \\ 
PSB Pair & 13.4 & 12.8  \\ 
Observation & 22.6 & 45.0  
\enddata 
\label{diffmaptable}
\end{deluxetable*} 

\begin{figure}
\centering
\includegraphics[scale=0.33]{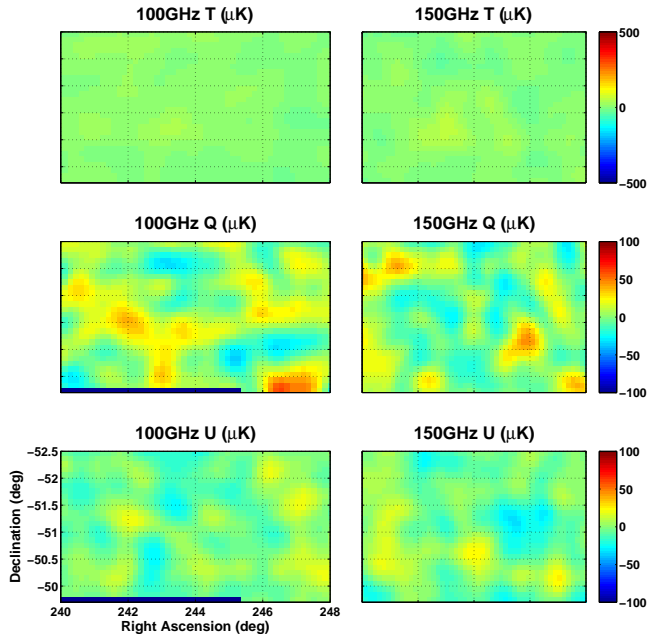}
\caption{T, Q, and U differenced data from individual detectors within a single PSB pair. The Q and U scales are identical to Figure 7 for comparison, whereas the temperature scale is reduced to $\pm500\,\mu K$. Large scale features in both polarization and temperature are no longer apparent in the maps.}
\label{pixeljack}
\end{figure}

Three separate divisions of the data were analyzed: scan-direction, PSB pair, and observation.  The scan-direction split, where the two data sets are separated based on half-scan direction, can generate noise based on thermal instability at half-scan endpoints. The scan-direction difference maps are shown in the last two rows of Figure~\ref{deepint}. PSB pair difference maps subtract the maps accumulated from individual detectors within a pair. This difference map is unique to FRM analysis, as a FRM turns a single BICEP PSB into a polarimeter, allowing T, Q, and U maps to be accumulated for individual detectors within a PSB pair. This difference map is perhaps the most robust data quality test for the FRMs as it is sensitive to many factors including relative gain mismatches, demodulation errors between detectors, and thermal stability. The PSB pair difference maps are shown in Figure~\ref{pixeljack}. Finally, observation difference maps subtract maps accumulated from odd and even numbered observing runs. This jackknife probes for sensitivity to weather changes. The FRM maps show the highest sensitivity to changes in weather, especially in the 150 GHz PSB.  The main source of the elevated signal at 150 GHz is a very high noise contribution at the lowest declination, where the least integration time occurs. 

\begin{figure}
\centering
\includegraphics[scale=0.33]{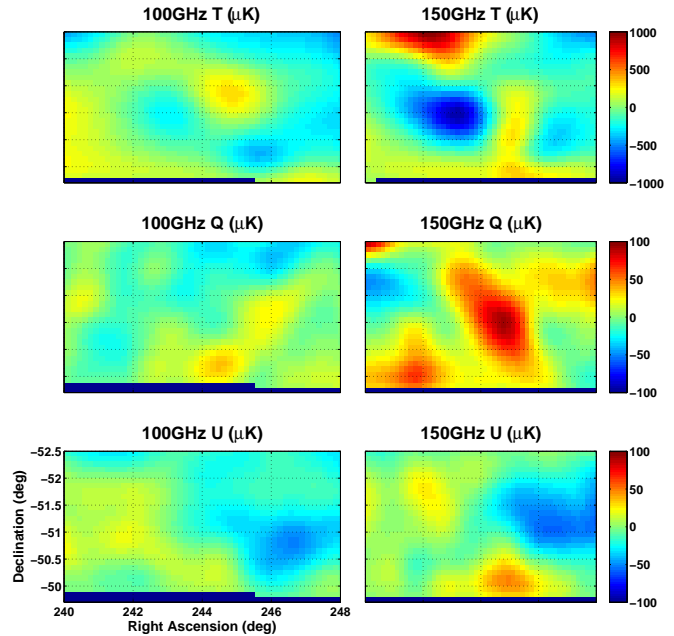}
\caption{Maps of the difference between the FRM and non-FRM BICEP observations. Galactic signal, in both temperature and polarization, has been removed. This is particularly evident in celestial U, where both the FRM and non-FRM BICEP maps originally exhibited significant polarized structure. No similar structure is evident in the Stokes U difference map.}
\label{BICEPvFRM}
\end{figure}

Difference maps of the two data sets, the FRM and non-FRM BICEP, are displayed in Figure~\ref{BICEPvFRM}. The rms map noise per pixel is given by 20.6 $\mu$K rms/pixel at 100 GHz and 35.6 $\mu$K rms/pixel at 150 GHz. This differencing should remove all true sky signal and leave only uncorrelated noise, which should combine as the quadrature sum of the noise in the two individual maps. To estimate the map noise in the non-FRM BICEP map, the scan direction difference map was utilized, yielding 12.5 and 12.0 $\mu$K per effective pixel at 100 and 150 GHz respectively. Given the map noise values in Table~\ref{pixeljack} for the FRM map, the difference map noise is consistent with the quadrature sum of the map noise of the FRM and non-FRM BICEP difference maps. 

Although many properties of the FRMs were characterized during initial testing and deployment, the lack of time for field testing prevented a complete characterization for the FRMs and is beyond the scope of this paper. Specifically, properties such as cross-polar leakage and polarization inefficiency were not studied in-depth, and it was assumed throughout this analysis that the FRMs behaved as ideal devices in these respects. Further, the DSC calibration revealed an imperfect bias signal for most of the FRMs, but for the deep observations it was assumed that the FRMs accomplished an ideal rotation of $\pm 45^{\circ}$ to the nominal angle of the coupled PSBs. Deviations from this perfect modulation would result in miscalibrated detector angles and reduced polarization efficiency, which would alter the polarization signal in the Q and U maps. 

Although the FRM maps are absolutely calibrated via non-FRM BICEP maps, the calibration is done using temperature alone. Deviations from the (assumed perfect) polarization properties of the FRMs (described above) would result in marked differences between the FRM and non-FRM Q and U difference maps. The fact that these maps are statistically consistent with each other limits the presence of FRM non-idealities to the percent level. 

\subsubsection{Map Noise}

\begin{figure}
\centering
\includegraphics[scale=0.35]{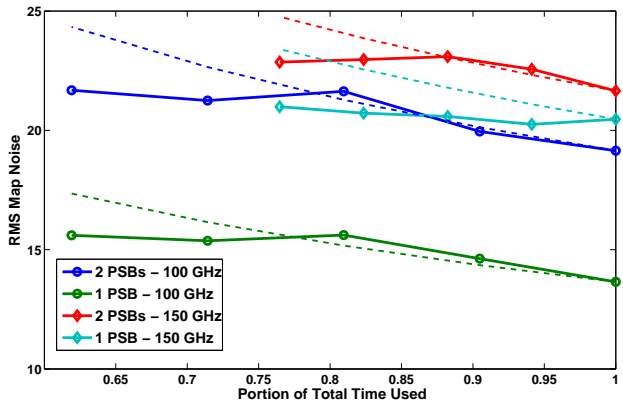}
\caption{Plot of rms map noise for the scan-direction difference maps as a function of the portion of total integration time used. At each frequency, two lines are displayed representing the rms difference map noise from individual PSB and the PSB pair. The thinner dashed lines are a guide to the eye exhibiting square-root of integration time dependence; they are plotted in the same color as their corresponding solid lines. At 100 GHz, the square-root of integration time scaling is obvious, whereas for a single PSB at 150 GHz there is no such dependence. In addition, the 100 GHz PSB pair scales as the square-root of the number of detectors ($\sqrt{2}$) from the individual PSBs to the pair. At 150 GHz, the $\sqrt{2}$ reduction in noise is not seen.}
\label{diffmap_nt}
\end{figure}

Map noise in FRM observations 2--5 was quantified via the scan-direction difference map. If the map was not sensitive to systematic effects (e.g. - systematics are not present), then the noise should be Gaussian white and integrate down with the square-root of the integration time and the square-root of the number of detectors. To validate this assumption for these maps, the data were reprocessed several times, each time removing a different number of points from the transitions during demodulation. Maps were accumulated using a range of cuts from four points (two on either side of the transition) to twelve points (six on either side of the transition). The results are shown in Figure~\ref{diffmap_nt}. The data exhibit a decrease in rms map noise with a $\sqrt{t}$ dependence based on the data cuts. Additionally, at 100 GHz, the data show a $\sqrt{2}$ increase in the map noise when accumulating the difference maps with a single PSB as opposed to the pair. At 150 GHz, however, the $\sqrt{2}$ dependence on PSB number is not observed.

\begin{figure}
\centering
\includegraphics[scale=0.46]{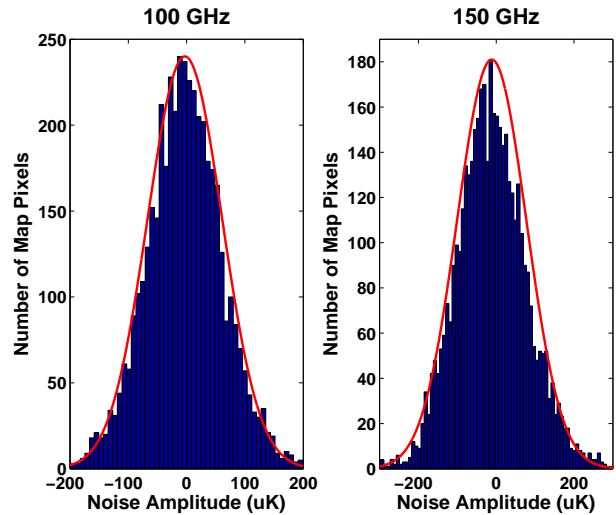}
\caption{Histograms of the noise at 100 GHz (left) and 150 GHz (right) in the unsmoothed scan direction jackknife maps for the FRM deep observations. Gaussian fits using the derived mean and standard deviation at each frequency are shown for comparison. }
\label{noisehist}
\end{figure}

Histograms of the noise distribution were also computed using the unsmoothed difference maps at both frequencies and are shown in Figure~\ref{noisehist}. Only map pixels with an integration time of 50 seconds or more are used in this analysis to avoid pixels with low integration time weighting the result. The histograms show that the amplitude distribution of the noise is roughly Gaussian and distributed about zero within one standard deviation of the mean. The mean $\pm$ standard deviation are given by $-4.1 \pm 75.3 \, \mu K$ and $-11.1 \pm 99.4 \, \mu K$ at 100 GHz and 150 GHz respectively. 

Assuming these maps are Gaussian white noise-dominated, detector noise was quantified by multiplying the Q and U pixel values by the square-root of the integration time per pixel and taking the standard deviation across the maps. This yielded an average instantaneous (i.e., single Stokes parameter) "NEQ per feed" of 420 $\mu K \sqrt{s}$ and 501 $\mu K \sqrt{s}$ for 100 GHz and 150 GHz respectively. The value at 100 GHz is comparable to the values in \citep{Takahashi09, Bierman11}, though the value at 150 GHz is elevated by a factor of ~1.6. The source of the excess noise at 150 GHz is unknown, though the lack of $\sqrt{2}$ dependence when doubling the number of PSBs indicates that systematic effects, such as detector cross-talk, can not be ruled out. 

\section{Results from Unbiased FRMs}

For the majority of the 2006 observing season, the FRMs remained unbiased in the BICEP focal plane. Without modulation, the PSB pairs coupled to the FRMs functioned as nominal BICEP PSBs, where the sum and difference of the pair were used to calculate temperature and polarization respectively. Using 2006 Galactic scans, noise was computed for all FRM feeds to analyze whether the presence of the FRMs in the optical path introduced additional noise within the detectors. 

\begin{figure}
\centering
\includegraphics[scale=0.5]{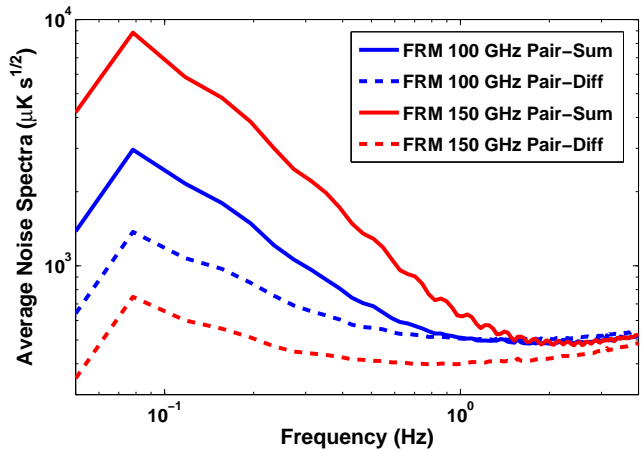}
\caption{Average power spectral distributions calculated for the pair-sum and pair-differenced FRM PSBs at 100 and 150 GHz during Galactic observations in 2006. Data for the plots were taken while the FRMs were unbiased in the BICEP focal plane. The second order polynomial filtering can be seen at 0.05 Hz, where the pair-difference noise is lower than the white noise floor.}
\label{noisespectra}
\end{figure}

To compute the noise, the spectral power distributions for detector sums
and differences for each calibrated half-scan is calculated and averaged for all FRM pixels at each band. This is done after masking, filtering, and sum-differencing each PSB pair. Figure~\ref{noisespectra} shows the results of this analysis. The pair-sum data exhibits increasing 1/$f$ contamination from 100 to 150 GHz. Pair differencing removes this contamination, resulting in nearly white noise above 0.1 Hz. Averaging the pair-difference periodogram between 0.1 and 1.0 Hz gives a Noise Equivalent Temperature (NET) per detector of 620 and 430 $\mu K\sqrt{s}$ for FRM feeds at 100 and 150 GHz respectively. Previous studies \citep{Takahashi09, Bierman11} characterizing the pixels in BICEP (which do not include the FRMs) give BICEP NET values of 525 $\mu K\sqrt{s}$ at 100 GHz and 450 $\mu K\sqrt{s}$ at 150 GHz. The elevated noise at 100 GHz was found to stem from the PSB pair corresponding to a single FRM (100C). When removed, the FRM NET at 100 GHz is 520 $\mu K\sqrt{s}$. Both bands show excellent agreement with the quantities derived from observation quality pixels, demonstrating that unbiased FRMs do not introduce any excess noise into the detectors. 

\section{Conclusions}

In this paper, we report on the first detection of Galactic polarization using fast, active optical polarization modulation. The observed polarized signal is consistent with both BICEP's and other recent experiments' maps of Galactic polarization. Although observation and testing time limited the scope of the comparison that could be made between the nominal BICEP instrument and the BICEP instrument modulated by FRMs, the FRMs have been shown to be effective solid-state polarization modulators for both laboratory and celestial polarimetry applications. In particular, the 100 GHz FRM was found to exhibit equivalent noise to similar BICEP PSB pairs without FRMs, but expands the parameter space for the instrument's scan strategy, relaxing constraints on scan speed and potentially increasing the observed sky fraction. Ultimately, the choice of whether to use a fast active modulator such as a FRM depends upon the details and constraints of the experiment. Devices similar to the FRMs are currently being explored for use in several experiments \citep{Tartari09, Gault12}. FRM technology could be applied wherever fast solid-state polarization modulation is called for.

FRMs are a promising technology due to their design flexibility, low systematic polarization, large bandwidth, and ability to be used over a large frequency range. Further, FRMs have been shown to be functional in both modulating and non-modulating modes without adding noise, making them a flexible option for a mm-wave polarization modulation. FRMs also have potential application as modulators for cm and sub-mm wavelength polarimeters.

\section{Acknowledgments}

S.M. gratefully acknowledges the American Association of University Women for a fellowship supporting this research. B.G.K also acknowledges NSF PECASE Award No. AST-0548262. G.R. gratefully acknowledges support by the NASA Science Mission Directorate via the US
Planck project.

BICEP is supported by NSF Grant OPP-0230438, Caltech Discovery Fund, Caltech President’s Fund PF-471, JPL Research and Technology Fund, and the late J. Robinson.

\end{document}